\documentclass[12pt,a4paper]{article}
\usepackage{graphics}
\usepackage{amsfonts}

\newcommand{\be}{\begin{equation}}
\newcommand{\ee}{\end{equation}}
\newcommand{\eel}[1]{\label{#1}\end{equation}}
\newcommand{\bea}{\begin{eqnarray}}
\newcommand{\eea}{\end{eqnarray}}
\newcommand{\eeal}[1]{\label{#1}\end{eqnarray}}
\newcommand{\baq}{\begin{equation}\begin{array}{rcl}}
\newcommand{\eaq}{\end{array}\end{equation}}
\newcommand{\eaql}[1]{\end{array}\label{#1}\end{equation}}
\newcommand{\beac}{\begin{equation}\begin{array}{rcl}}
\newcommand{\eeacn}[1]{\end{array}\label{#1}\end{equation}}
\newcommand{\ba}{\begin{array}}
\newcommand{\ea}{\end{array}}
\newcommand{\non}{\nonumber \\}
\newcommand{\equ}[1]{(\ref{#1})}

\newcommand{\beq}{\begin{eqnarray}}
\newcommand{\eeq}{\end{eqnarray}}
\newcommand{\nn}{\nonumber}

\newcommand{\journal}[4]{{\rm #1~}{\bf #2}\,(19#3)\,#4}

\newcommand{\np}{\journal {Nucl. Phys.}}
\newcommand{\pl}{\journal {Phys. Lett.}}

\begin{document}

\thispagestyle{empty}

\begin{titlepage}

\begin{flushright}
\textsf{TAUP--2433--97, UTTG--20--97}\\
\textsf{hep--th/9706127}\\
\textsf{June 1997}
\end{flushright}
\vfill
\begin{center}
\textbf{\large Comments on the M Theory Approach to}\\
\textbf{\large N=1 SQCD and Brane Dynamics} 
\vskip 1cm
\textsc{A. Brandhuber$^a$, N. Itzhaki$^a$, V. Kaplunovsky$^{b,1}$,}\\
\textsc{J. Sonnenschein$^{a,}$\footnote{Research supported in part by: 
NATO, R. A. Welch Foundation, NSF grant PHY9511632 (V.K.);
the US-Israel Binational Science Foundation (V.K, J.S. \& S.Y.);
the German-Israeli Foundation for Scientific Research ( J.S. \& S.Y.);
the Israel Science Foundation (J.S. \& S.Y.).}
and S. Yankielowicz$^{a,1}$}\\[5mm]
\emph{$^a$School of Physics and Astronomy}\\
\emph{Beverly and Raymond-Sackler Faculty of Exact Sciences}\\
\emph{Tel-Aviv University, Ramat-Aviv, Tel-Aviv 69978, Israel}\\[3mm]
\emph{and}\\[3mm]
\emph{$^b$Theory Group, Dept. of Physics, University of Texas}\\
\emph{Austin, TX 78712, USA}
\end{center}
\vfill

\begin{abstract}
We use the M theory approach of Witten to investigate
N=1 SU($N_c$) SQCD with $N_f$ flavors.
We reproduce the field theoretical results and identify
in M theory the gluino condensate and the eigenvalues
of the meson matrix.
This approach allows us to identify the constant piece 
of the effective field
theory coupling from which the coefficient of the 
one-loop $\beta$-function can
be identified. 
By studying the area  of the M-theory five-brane 
we  investigate the stability of  type IIA brane
configurations. We prove that in a supersymmetric setup
there is no force between static D4-branes that end on
NS five-branes.
The force in the case that there is a relative velocity between the
branes is computed. We show that at the regions of intersecting IIA
branes  the curvature of the M theory five-brane is singular in the
type IIA limit. 
\end{abstract}

\end{titlepage}

\setcounter{page}{1}

\section{Introduction}

By now it is clear that D-branes of string theory are an important
tool in constructing and unraveling the non-perturbative physics of
supersymmetric gauge theories. These gauge theories live on the world
volume of the corresponding D-brane. The interplay between the brane
dynamics and the gauge theory dynamics has proven to give important
results and insight into the non-perturbative physics of both of
them. Following the work of Hanany and Witten \cite{han} who
considered a particularly useful construction giving rise to N=4
supersymmetric theories in 2+1 dimensions many other works generalized
it to other dimensions \cite{egk,ah,bk,kol} and three dimensional
theories with four supercharges \cite{boer}.
In particular, the construction was
generalized to four-dimensional N=1 SQCD in \cite{egk} where also the
brane origin of Seiberg's duality  was revealed in \cite{sei1}. 
This approach was further
explored in refs. \cite{barbon,bsty1,bh,egkrs,hz} and generalized to
all classical gauge groups \cite{egkrs,ejs,tatar}. The same
class of field theories can also be studied by wrapping D branes on
cycles of Calabi-Yau manifolds \cite{v1,v2,v3,v4,ahn1,ahn2,at}. 

The basic set-up of the brane configuration giving rise to
supersymmetric SU($N_c$) SQCD involves two NS fivebranes with $N_c$ D4
branes stretched between them. Having one of the dimensions bounded
between the NS fivebranes the world-volume of the D4 branes is
effectively 3+1 dimensional. One can also include D6 branes to account
for matter degrees of freedom (flavors). Alternatively, the flavors
can be introduced via semi-infinite D4 branes emanating from the two
NS branes. 

In a recent paper Witten \cite{witten} has given an M theory
interpretation for the brane configuration associated with N=2
SU($N_c$) supersymmetric gauge theories in four dimensions. 
The NS fivebranes and the D4 branes are 
generated in the $R_{10} \to 0$ limit, in which type IIA theory is
recovered, from a single M theory fivebrane (M5). The world volume of
the M5 brane is $\mathbb{R}^{1,3} \times \Sigma$, where $\Sigma$ is
embedded holomorphically in a $\mathbb{R}^3 \times S^1$ part of eleven
dimensions. To incorporate D6 branes $\mathbb{R}^3 \times S^1$ has to be
replaced by a multi-Taub-NUT space. This approach provides a
geometrical picture of the corresponding four-dimensional
supersymmetric quantum gauge theory. In particular the beta function
in the case of N=2 theories acquires a geometrical interpretation.
This construction has been generalized to all classical gauge
groups in \cite{lll,bsty2}. This approach has also been studied in
\cite{mmm}. 

In the present paper, by analyzing the M theory five-brane we derive known
non-perturbative results of 4d N=1 SQCD and investigate the stability
of the type IIA brane configurations that admit 4d N=2 and N=1
supersymmetric gauge theories.

Section 2 will be devoted to the M theory approach to N=1 SU($N_c$)
SQCD with $N_f$ flavors. In our formulation the matter will be
generically massive and represented by $N_f$ semi-infinite D4
branes. We shall show that the M theory approach correctly reproduces
the known non-perturbative field theoretical results associated with
the structure of the vacuum. In particular, we will identify within
the M theory approach the parameters associated with the gluino
condensate and with the meson field matrix eigenvalues. We will
establish relations between these parameters and the masses of the
quark hypermultiplets which were previously derived within the
effective potential approach by Taylor, Veneziano and 
Yankielowicz \cite{tvy}.
In section 3 we will
discuss the stability of the brane configurations. We shall prove a
general theorem showing that as long as we are dealing with $\Sigma$
which is holomorphically embedded in $\mathbb{R}^{2n}$
 (or 
$\mathbb{R}^{2n-d} \times T^d$ ) any parameter 
which may appear is a real
modulus i.e. corresponds to a flat direction of the
superpotential. Thus the classical potential which corresponds to the
minimal area cannot depend on it. This will amount to no static force. 
As an example we will demonstrate this general result in one case
(relevant to $N=2$ or $N=1$ supersymmetric gauge theories) involving
D4-branes and NS fivebranes.
In section 4 we  show that between D4-branes with relative velocity
$v$ there is a force that depends on $v^2$. In the field theoretical
language it corresponds to evaluating the K\"ahler function
(while the static force is related to the superpotential).
Another feature of the metric on M5 that will be investigated in
section 5  is the curvature. We will show that  at the ``point" where
D4-branes end on NS five-branes there is a curvature singularity  in
the type IIA limit. 

While finishing this paper we have received a preprint by Hori, Ooguri
and Oz \cite{hoo} who consider N=1 SQCD as a flow from N=2 SQCD. They
also introduced D6 branes and considered the massless as well as the
massive case. Furthermore, a preprint by Witten appeared
\cite{witten2} which applies the M theory approach to discuss chiral
symmetry breaking, confinement and domain walls in N=1 SYM.

\section{M-theory description of N=1 SQCD}

Pure N=1 supersymmetric Yang Mills theory is described by a brane
configuration involving NS, NS' and $N_c$ D4 branes stretched between
them. In the original set-up of \cite{egk} the world volumes
associated with each of the branes are
$(x_0, x_1, x_2, x_3, x_4, x_5)$ for the NS branes,
$(x_0, x_1, x_2, x_3, x_8, x_9)$ for the NS' branes, and
$(x_0, x_1, x_2, x_3, x_6)$ for the D4 branes.

The 4d space-time  $\mathbb{R}^{1,3}$ with coordinates $(x_0, x_1, x_2,
x_3)$ will not be important in what follows and we will ``ignore''
it.  The world-volume of the M theory five-brane is taken to be
$\mathbb{R}^{1,3} \times \Sigma$ \cite{witten}, where $\Sigma$ is a complex
Riemann surface. A 4d  $N=2$ supersymmetry was shown to be associated
with the embedding of $\Sigma$ in $Q \simeq \mathbb{R}^3 \times S_1$ which
is equipped with a  complex structure in terms of
\be
v=x_4 +i x_5~,~ t = \exp \left( - (x_6 + i x_{10})/R_{10} \right),
\ee
where $x_{10}$ is the  coordinate along  the eleventh compactified
dimension $S_1$ of radius $R_{10}$ and the remaining coordinates $x_7$,
$x_8$ and $x_9$ are held constant. 

The M theory five-brane  setup 
which admits  only $N=1$ supersymmetry in $\mathbb{R}^{1,3}$ requires
the embedding 
of $\Sigma$ in $Q \simeq \mathbb{R}^5 \times S_1$   with a global 
complex structure in terms of $v$, $t$ and  $u = x_8 + i x_9$. The latter
condition implies that $v(u), u(t)$ and $v(t)$ are all
holomorphic functions.  

We start with the classical configuration where the NS and the NS'
brane are orthogonal in the $(x_4, x_5, x_8, x_9)$ space. The
coordinate $v$ is associated with the NS fivebrane and the coordinate
$u$ with the NS' fivebrane. 
In M-theory terms, quantum effects are manifested via bending of all the
branes. Thus, the NS brane spans a finite range of $u$ while the NS'
spans a finite range of $v$. However, only the NS  brane extends  into
the asymptotic $v\rightarrow\infty$ region. Therefore, the holomorphic
functions $u(v)$ and $t(v)$ are single valued for large $v$ which in
turn implies single valuedness of $u(v)$ and $t(v)$ throughout the
complex $v$ plane. Similarly, only the NS' brane  exists in the
asymptotic $u\rightarrow\infty$ region and hence the functions $v(u)$
and $t(u)$ should also be single valued. In particular, the holomorphic map between $v$ and $u$ is bijective; all such maps are of the form
$u = (a v + b)/(c v + d)$ with  $a, b,
c, d \; \epsilon \; \mathbb{C}$ and $ b c \neq 0$ to ensure the correct
asymptotic behavior.
Hence, after suitable constant shifts of the complex $u$ and $v$
coordinates, the embedding of the M theory fivebrane into
$\mathbb{R}^4_{4,5,8,9}$ is described as
\be
u v = S = \textrm{const}~.
\eel{eq4}

In the pure SU($N_c$) SYM theory the NS fivebrane has $N_c$ D4 branes
attached to its left side (in the $x_6$ direction) while the NS'
fivebrane has $N_c$ D4 branes attached to its right side. Hence,
according to Witten \cite{witten} 
\be
t-P(v) = 0\;,~\ \ \  Q(u) t - A = 0
\eel{eq6}
where $A$ is a normalization constant and
\beq\label{new}
P(v)=v^{N_c}+p_1 v^{N_c -1}+...+p_{N_c},\\ \nn
Q(v)=u^{N_c}+q_1 u^{N_c -1}+...+q_{N_c},
\eeq
are some polynomials of degree $N_c$.
Note that both equations \equ{eq6} are linear in $t$ in accordance with the single-valuedness of the holomorphic functions $t(v)$ and $t(u)$.

Consistency of eqs. \equ{eq4} and \equ{eq6} requires 
\be
P(v) Q(u = S/v) \equiv A,\qquad \forall v\in \mathbb{C}
\eel{eq7}
Note that this is not an equation for $v$ (that would select just a
point on the curve $\Sigma$) but a set  of conditions for the
parameters $p_1,...,p_{N_c}$, $q_1,...,q_{N_c}$ and $A$ that would
make \equ{eq7} an identity with respect to $v$.
The unique solution of this system is
\be
P(v) = v^{N_c},\;\; Q(u) = u^{N_c},\;\; S^{N_c} = A ,
\ee
which tells us that there are $N_c$ discrete vacua related by a
$\mathbb{Z}_{N_c}$  symmetry -- a well known result in SYM
\cite{index,vy}. In all these vacua the $N_c$ D4 branes sit on top of
each other.

Next, we would like to add matter to the SYM theory. 
In terms of the string/M-theory setup, this can be done by  attaching
semi-infinite D4-branes to the NS or NS' branes and/or inserting
D6-branes between NS and NS'. We are interested in massive quarks with
generic masses $m_1,...,m_{N_f}$ and use the simpler semi-infinite
D4-brane setup; the massless case is more involved and is (presumably)
more tractable in the D6-brane setup. 

Without loss of generality, we attach  all the semi-infinite D4-branes
to the same NS brane. 
Thus, following ref. \cite{witten}, we replace eq. \equ{eq6} with
\bea
R(v) t - P(v) &=& 0, \non
Q(u) t - A    &=& 0.
\eeal{eq10}
where $P(v)$ and $Q(v)$ are polynomials of degree $N_c$,
(cf. eq. \equ{new}) while 
\be 
R(v)=\prod_{i=1}^{N_f} (v-m_i).
\eel{eq8}
In particular, for $N_f=0$, $R(v)=1$ and we recover eq. \equ{eq6}.

By a suitable rescaling of the variable $t$ we can always set the
leading terms of the polynomials $P$, $Q$ and $R$ to $v^{N_c}$,
$u^{N_c}$ and $v^{N_f}$ respectively. This leaves us with one
normalization factor $A$, which we shall now identify with
$\Lambda_{QCD}^{3 N_c - N_f}$ for the following reasons:
Consider the geometric symmetry $U(1)_{45} \otimes U(1)_{89}$ of the M
theory associated with complex rotations of the NS and NS'
fivebranes. The charges of the coordinates $v$, $u$ and $t$ and
various parameters of the curve $\Sigma$ are summarized in Table 1
\begin{table}
\begin{center}
\begin{tabular}{|r|c|c|c|c|c|c|c|c|}
\hline
charge      & $v$ & $u$ & $S$ & $t$ & $A$ & $m_i$ & roots of $P(v)$ &
roots of $Q(u)$ \\
U$(1)_{45}$ & 1 & 0 & 1 & $N_c - N_f$ & $N_c - N_f$ & 1 & 1 & 0 \\
U$(1)_{89}$ & 0 & 1 & 1 & 0           & $N_c$       & 0 & 0 & 1 \\
\hline
\end{tabular}
\end{center}
\caption{}
\end{table}
(The charges of $u$ and $v$ are obvious, the rest follows from
eqs. \equ{eq4}, \equ{eq10} and \equ{eq8}.)

Now consider the SQCD, which has two broken abelian symmetries
$U(1)_{\mathrm{A}} \otimes U(1)_{\mathrm{R}}$ (one linear combination
of $U(1)_{\mathrm{A}}$ and $U(1)_{\mathrm{R}}$ is broken by the quark
masses while the other is anomalous). Table 2 below summarizes the
charges of various physical quantities under these symmetries, as well
as under two linear combinations $U(1)_\phi$ and $U(1)_\psi$ leaving
invariant the squarks $\phi_L$ and the quarks $\psi_L$, respectively.
\begin{table}
\begin{center}
\begin{tabular}{|r|l|l|l|l|l|l|l|}
\hline
charge & $\phi_L$ & $\psi_L$ & $\lambda$ & $S = \langle
\lambda \lambda \rangle$ & $M_{ij} =\langle \phi^L_i\phi^R_j\rangle$ & $m_i$ &
$\Lambda_{QCD}^{3 N_c - N_f}$ \\ 
$Q_{\mathrm{A}}$ & 1 & 1 & 0 & 0 & 2 &--2 & $2 N_f$ \\
$Q_{\mathrm{R}}$ & 0 &--1 & 1 & 2 & 0 & 2 & $2 N_c - 2 N_f$ \\
$Q_\phi = \frac{1}{2} Q_{\mathrm{R}}$ & 0 & $-\frac{1}{2}$ & $\frac{1}{2}$
& 1 & 0 & 1 & $N_c - N_f$ \\
$Q_\psi = \frac{1}{2} (Q_{\mathrm{A}} + Q_{\mathrm{R}})$ & $\frac{1}{2}$
& 0 & $\frac{1}{2}$ & 1 & 1 & 0 & $N_c$ \\
\hline
\end{tabular}
\end{center}
\caption{}
\end{table}
Note that $\Lambda_{QCD}^{3 N_c - N_f} \sim \exp (2 \pi i \tau) = -
\frac{8 \pi^2}{g^2} + i \theta$ is charged under anomalous symmetries;
indeed, its charge is precisely the anomaly of the symmetry.

A quick comparison of Tables 1 and 2 immediately suggests that the
$Q_\psi$ should be identified with $Q_{89}$ while $Q_\phi$ should be
identified with $Q_{45}$. Given this charge identification, we see
that the $A$ parameter of the M theory (cf. eq. \equ{eq10}) has 
the same charge as $\Lambda_{QCD}^{3 N_c - N_f}$ parameter of the
field theory.

Furthermore, $\log A$ is associated with a constant
(i.e. it is $u$ and $v$ -independent) term in the $x_6$ distance
between the NS and the NS' fivebranes. A constant shift in $x_6$
amounts to a real rescaling in $t$ which, according to eq. \equ{eq10},
amounts to a rescaling of $A$ and therefore a shift in $\log
A$. Similarly in SQCD there is also a $\log \Lambda_{\mathrm{SQCD}}^{3
  N_c - N_f}$ term in the field dependent effective coupling $8
\pi^2/g^2$. Since the $x_6$ distance between the NS and NS' fivebranes
corresponds to $8 \pi^2/g^2$ in the field theory, we have yet another
reason to identify 
\be
A = \Lambda_{\mathrm{SQCD}}^{3N_c - N_f}. 
\eel{eq12}

Equation \equ{eq10} implies
\be
t = \frac{P(v)}{R(v)} = \frac{A}{Q(u)}~
\eel{eq13}
and hence
\be
A R(v) = P(v) Q(u = S/v).
\eel{eq14}
Again, eq. \equ{eq14} should hold identically for every $v$, which
implies a system of
equations for $S$ and the coefficients of the polynomials $P_{N_c}(v)$
and $Q_{N_c}(u)$. (The coefficients of $R_{N_f}(v)$ are already
determined via the masses -- see eq. \equ{eq8}). The unique solution
is:
\bea
&&P=v^{N_c}, \non
&&Q=u^{N_c - N_f} \prod_{j=1}^{N_f} (u - S/m_j), \\
&&S^{N_c}=\Lambda^{3 N_c - N_f} \prod_{j=1}^{N_f} (-m_j).
\eeal{eq15}

By comparison, SQCD gives rise to an effective superpotential
\cite{tvy}
\be
W_{\mathrm{eff}}(S,M) = \mathrm{tr} \left( m M \right) + S \left(
  \log \left( \frac{S^{N_c - N_f} \mathrm{det} M}{\Lambda^{3 N_c -
        N_f}} \right) - (N_c - N_f) \right),
\ee
where $S = 16 \pi^2 \langle \mathrm{tr} \lambda^\alpha \lambda_\alpha
\rangle$ is the gaugino condensate and $M$ is the meson matrix
$\langle\phi_i \tilde{\phi}_j \rangle$ of the quark-anti-quark
condensates. When all the quarks are massive, SQCD has the same Witten
index as pure SYM, namely $N_c$ and hence should have $N_c$ discrete
vacua. Indeed, extremizing the superpotential (14) with respect
to $S$ and $M_{ij}$ we obtain 
\bea
S^{N_c} & = & \Lambda^{3 N_c - N_f} \mathrm{det} (-m) ,\non
M & = &- S m^{-1}.
\eea
(In the decoupling limit where all masses are large 
the first equation reproduces the well-known relation of N=1 pure SYM
$S^{N_c} = \Lambda_{eff}^{3N_c}\equiv
\Lambda^{3 N_c - N_f} \mathrm{det} (-m)$) 

Thus, $S$ of the M theory (eq. \equ{eq4}) is identified with the
gaugino condensate while the non-zero roots of $Q(u)$ are identified
as the eigenvalues of the ``meson" matrix.
The physical origin of this correspondence needs further investigation.

\section{Stability of brane configurations from M theory}
When a multi-brane configuration of string theory is used to elucidate
the vacuum structure of a field theory in four dimensions, the positions
of various branes in $x^4,\ldots,x^{10}$ should be static, {\it i.~e.},
independent on $x^0,\ldots,x^3$;
consistency of such a configuration requires static balance of forces
exerted by the branes upon each other.
In the M theory description, the web of multiple D4, NS, NS', {\it etc.},
branes becomes a single convoluted five-brane of the form
$\mathbb{R}^{1,3}_{0,1,2,3}\otimes \Sigma $ where $\Sigma$
is a two-dimensional surface embedded in the space spanned by the
$x^4,\ldots,x^{10}$.
(This space may be curved as the parallel D6 branes of the IIA theory
translate into a multi-Taub-NUT metric for the coordinates
$x^{4,5,6,10}$.) The potential energy of this fivebrane (or rather the
energy density in four dimensions) is simply the brane tension $T_5$
times the area of the surface $\Sigma$ and the requirement of the
static force balance means simply that that area --- as a function of
the shape of the $\Sigma$ --- should be at a local minimum.

For the sake of the four-dimensional supersymmetry, we would like $\Sigma$
to be a holomorphic complex curve in a K\"ahler space.
That is, the metric of the embedding space should be K\"ahler with respect
to three complex coordinates such as
\be\label{aaxxiii}
v=x^4+ix^5,\quad
u=x^8+ix^9,\quad
s=x^6+ix^{10}
\ee
and flat with respect to the remaining real coordinate $x^7$ while the
surface $\Sigma$ is spanned by
\be
x^7={\rm const},\quad v=v(z),\quad u=u(z),\quad  s=s(z)
\ee
for some holomorphic functions $v(z)$, $u(z)$ and $s(z)$ of the
$\Sigma$' coordinate $z$.
It is well known from the theory of holomorphic world-sheet instantons
that all such surfaces automatically satisfy the local minimal-area
variational equations.
In other words,
supersymmetry implies holomorphy which in turn implies that the
functions are automatically harmonic. 
But the converse is not true in general.
A solution satisfying the minimal-area equations can
correspond to a non-supersymmetric configuration. Examples of brane
configurations in type IIA string theory without supersymmetry can be found in
\cite{bsty1}, related non-supersymmetric brane configurations in M
theory have recently been studied in \cite{witten2}. 

Unfortunately, the surfaces we are interested in have non-compact
features describing the asymptotic NS branes and semi-infinite D4
branes and hence infinite total areas.
Consequently, the boundary contributions to possible finite variations
of the infinite area become a non-trivial problem of regularization
and  boundary conditions.

We are now going to show that a properly regularized area is independent
on any moduli of the surface $\Sigma$ that does not affect its asymptotic
behavior in any of the non-compact directions.
Such moduli --- for example, those describing the positions of finite
D4 branes connecting two parallel NS branes --- have truly flat potential
energies and the corresponding vacua of the four-dimensional field
theory remains exactly degenerate in the M theory.
In the IIA terms, even though the D4 branes bend the NS branes upon which they
terminate, the resulting net force between the D4 branes remains exactly zero.

To understand the proper regularization procedure for the infinite or
semi-infinite asymptotic regions of $\Sigma$, consider a classical mechanical
system involving a semi-infinite string under tension.
In order to correctly account for the force of the string's tension, the
infinite potential energy of the string should be regularized by replacing
the infinite terminus of the string with one at {\em fixed location} at some
finite but very large distance from the other end of the string.
Likewise, the potential energy of an infinite membrane with tension should
be regularized by considering a large but finite membrane attached to a fixed
boundary very far away.

Clearly, the same prescription applies to the $\Sigma$ surfaces of the M
theory: A tube describing a semi-infinite D4 brane extended to
 $x^6\to\pm\infty$ should be terminated at large but fixed
$|s+\bar s|=\lambda_s$ while an asymptotic NS brane extending to
 $x^{4,5}\to\infty$ should be terminated at large but fixed $|v|=\lambda_v$;
in $N=1$ configurations, the NS' branes should be likewise terminated
at a large fixed $|u|=\lambda_u$.
Note that this cutoff does not violate the $U(1)_{45}\times U(1)_{89}$
symmetry of the NS and NS' or the cyclical nature of the $x^{10}$ coordinate.

As an example, consider an $SU(N_c)^K$ theory with $N=2$ SUSY of
ref.\cite{witten}; for simplicity we take same $N_c$ for all the gauge
groups. The brane setup consists of $K+1$ parallel NS branes connected
by D4 branes; there are $N$ D4 segments between each pair of adjacent
NS branes and also $N$ semi-infinite D4 branes attached to the first
and to the last NS branes; there are no D6 branes.
In the M theory, the embedding space is flat $\mathbb{R}^6\times S^1$ and the
surface $\Sigma$ is described by
\be\label{bigsurface}
u=0,\qquad
t^{K+1}P^{(K+1)}(v)+t^KP^{(K)}(v)+\cdots+t^0P^{(0)}(v)=0
\ee
where $t=\exp(-s/R_{10})$ and
each of the $P^{(K+1)}(v),\ldots,p^{(0)}(v)$ is a polynomial in $v$
of degree $N_c$;
their coefficients are moduli of the $\Sigma$ encoding the positions
of all the NS branes and the D4 brane segments.
Since the embedding space is flat, the induced metric on $\Sigma$ is
\be\label{inducedmetric}
h_{z\bar z}dzd\bar z = dv(z)d\bar v(\bar z) + ds(z)d\bar v(\bar z)
\ee
and hence the area of $\Sigma$ is simply
\be\label{surfaceareaint}
\int\!\!\int\! dzd\bar z\,h_{z\bar z}(z,\bar z)\
= \int\!\!\int\! dvd\bar v\ + \int\!\!\int\! dsd\bar s
\ee
and the only question is the precise  region of integration.

In the absence of regularization, $\Sigma$ covers the complex $v$ plane
$K+1$ times and the complex $t$ plane $N_c$ times, which corresponds
to a $N_c$-fold cover of the cylinder spanned by the $s$ coordinate.
The regularization of $\Sigma$ results in an $(K+1)$-fold cover of the
radius-$\lambda_v$ circle in the $v$ plane with a few tiny holes cut out where
$s+\bar s$ goes to $\pm\infty$ {\it i.~e.}, $t\to0$ or $t\to\infty$.
Such holes are located at zeros of the polynomials $P^{(0)}(v)$
and $P^{(K+1)}(v)$ and have sizes of the order $O(e^{-\lambda_s/2R_{10}})$.
Although the exact sizes and locations of such holes depend on the
moduli of the curve, the holes are so small that they can be simply
ignored altogether while performing the $\smallint\!\smallint dvd\bar
v$ integral. As the result, for large $\lambda_s$, the first integral
on the right hand side of eq.~\ref{surfaceareaint} becomes
$\pi(K+1)\lambda_v^2$ independent on any moduli of the surface.
Similarly, for the $s$ coordinate we have an $N_c$ cover of the cylinder
of radius $R_{10}$ and fixed height $\lambda_s$ with $K+1$ small holes
where $v\to\infty$. Again, the sizes and locations of the holes depend
on the moduli of $\Sigma$, but in the large $\lambda_v$ limit the
holes are negligibly small regardless of the moduli and the second
integral on the right hand side of eq.~\ref{surfaceareaint} become $2\pi N_cR_{10}\lambda_s$ independent on any moduli.

More complicated brane configurations and corresponding 
$\Sigma$ surfaces
can be analyzed in a similar way, with similar results.
The conclusion is that the area of properly regularized $\Sigma$ is indeed
independent on any moduli, the corresponding vacua of the field theory
remain exactly degenerate and there are no net forces between the D4
branes of the IIA theory.

\section{Dynamical force between moving branes}
In this section we use M theory configurations to
study the dependence of the force on the relative
velocity.
Forces between slowly moving infinite 
D-branes were studied in details in \cite{bac,lif,dkps}.
The force is proportional to $v_{\mathrm{rel}}^4$
(where $v_{\mathrm{rel}}$ is the relative velocity) when preserving 
sixteen of the super-charges and proportional to $v_{\mathrm{rel}}^2$ when preserving 
eight of the super-charges. 
The configurations that we consider in this section 
preserve eight of the super-charges. 
Therefore, the force should be proportional to $v_{\mathrm{rel}}^2$.

\subsection{Strings on D2-branes}

To simplify the discussion we start with an analogous type IIA
configuration of parallel D2 branes joined by fundamental
strings. This setup is described in M theory by a single M2 brane. The
curve  of such a configuration is identical to the curve of the
corresponding configuration of fourbranes between parallel fivebranes.
 
Our starting point is 
the Lagrangian of the M2 brane 
\be 
{\cal L} =-T_2\sqrt{-\det h},
\ee
where $h_{\alpha\beta}=\partial_{\alpha}x^{\mu}\partial_{\beta}x^{\nu}
g_{\mu\nu}$, $\alpha , \beta =0, 1, 2$.
Choosing $\xi_0=x_0$
the Hamiltonian  is
\be
\label{uuuu} H =T_2\int d \xi_1 d\xi_2 \frac{\partial_1 X^{\mu}\partial_1 X_{\mu}
\partial_2 X^{\nu}\partial_2 X_{\nu}-(\partial_1 X^{\nu}
\partial_2 X_{\nu})^2}{\sqrt{-\det h}}.
\ee  

The first case that we wish to study is a configuration of
two strings attached to one D2-brane from opposite sides.
The relevant curve is 
\be
\label{lklk} t(v-c/2)-v-c/2=0,
\ee
where $ c(x_0)=c_0 + v_{\mathrm{rel}} x_0$.
Thus the relative distance between the fourbranes
is $c$ and the relative velocity is $v_{\mathrm{rel}}$.
To first order in $v_{\mathrm{rel}}^2$ and $R_{10}^2/c^2$ one gets
from eqs.(\ref{uuuu}, \ref{lklk})
\be 
\label{uxu}H=-\frac{\pi}{2} T_2 v_{\mathrm{rel}}^2 R_{10}^2 \log (\lambda_v^2
 \lambda_t /c^2),
\ee
where $\lambda_t$ is the cutoff at the $t$ plane,
$1/\lambda_t<|t|<\lambda_t$, and $\lambda_v$ is the cutoff in the
$v$-plane, $|v|<\lambda_v$. The force between the ending points of the
fourbranes is attractive
\be 
F=\pi\frac{v_{\mathrm{rel}}^2 R^2_{10}T_2}{c}.
\ee

Now, let us consider two strings which end on the same
side of a 2D-brane. The curve is given by
\be
\label{rere} t-\frac{v^2-c^{2}/4}{a^2}=0,
\ee
where again $ c(x_0)=c_0 + v_{\mathrm{rel}} x_0$.
To first order in $v_{\mathrm{rel}}^2$ and $R_{10}^2/c^2$ one finds 
\be
\label{uu} H=-\frac{\pi}{2}
T_2 v_{\mathrm{rel}}^2 R_{10}^2 \log (\frac{\lambda_t c^2}{a^2}).
\ee
Note that $\lambda_v$ does not appear in this
 expression unlike in 
eq.(\ref{uxu}). The derivative with respect to $c$ gives  a repulsive
force
\be F=-\pi\frac{v_{\mathrm{rel}}^2 R^2_{10}T_2}{c}.\ee  

\subsection{D4-branes on NS fivebranes}

The only difference between a configuration of 
fourbranes ending on fivebranes and the corresponding 
configuration of strings ending on D2-branes is that the 
ends of the fourbranes depend on  
$x_0, x_1, x_2$ and $x_3$.
The generalization  of the result of the previous subsection
yields when the two fourbranes are on the same side of the
fivebrane 
\be
\label{9} H=-2\pi T_5 R_{10}^2
\int d^3 x(\partial_i c)^2\log
(\frac{\lambda_t c^2}{a^2}),
\ee
where $i=0, 1, 2, 3$ and $c=c(x_0, x_1, x_2, x_3)$.
Since $c$ plays the role of the expectation value of the
scalar field in the
four-dimensional theory this Hamiltonian should be
related to the kinetic energy in the four-dimensional theory.
To make this relation more precise we need to consider 
two NS fivebranes joined by two D4-branes. The relevant curve is 
\be 
t^2+\frac{v^2-c^{2}/4}{a^2}t+1=0,
\ee
where $c$ is the distance between the fourbranes.
When $c^2/a^2\gg 1$  the distance between the fivebranes
at $v=0$ is
\be
\label{cxcx} \Delta x_6=4R_{10}\log
(\frac{c}{a}).
\ee
This distance is related to the coupling of the gauge
theory on the four branes 
\be 
\frac{R_{10}}{g^2}=\Delta x_6.
\ee
When $c/a\rightarrow \infty$ we recover at the semi-infinite
result. This means that  $\lambda_t$ (which cannot appear in the
absence of semi-infinite 4-branes) should be replaced with $c^2/a^2$
(as eq.(\ref{cxcx}) implies). Hence in this limit eq.(\ref{9}) yields 
\be 
H\propto\frac{1}{g^2}\int d^3 x
(\partial_i c)^2.
\ee
As expected the computation of the dynamical force gives a first order
approximation to the (non-holomorphic) K$\ddot{\mbox{a}}$hler
function.
We recall that the static force is connected to the holomorphic
superpotential.
Clearly, our general discussion in section 3 relies heavily on
holomorphicity and does not apply to the K$\ddot{\mbox{a}}$hler
function.
\section{ The curvature}

The curve of the  M theory five-brane is smooth and invariant
under rescaling of $R_{10}$. Thus, the curve is not a useful tool 
to pass to  the type IIA brane configurations. Instead the
embedding metric on the five-brane, and in particular the
corresponding curvature, can be used to approach the non smooth
nature of the intersections of branes.
In this section we show that the ends of the D4-branes on the NS
fivebranes form curvature singularities in the type IIA limit. 

We begin with the simplest example, one D4-brane ending on one NS brane.
The corresponding curve is 
\be 
v t=1. 
\ee
The induced metric is 
\be 
ds^2=dvd\bar{v} (1+\frac{R_{10}^2}{\mid v\mid ^2}),
\ee
and  the curvature is given by 
\be 
R=2\frac{\mid v\mid ^2 R_{10}^2}
{(\mid v\mid ^2+R_{10}^2)^3}.
\ee

The curvature vanishes far on the NS brane ( $v\rightarrow \infty$)
and far on the D4-brane ($v\rightarrow 0$). The maximal value of the
curvature is obtained in the region where $\mid v\mid ^2\approx R_{10}^2$
\be R_{max}\approx \frac{1}{R_{10}^2}.\ee

Let us  consider now two D4-branes attached to one NS brane.
As long as the distance between them is larger then
$R_{10}$ we expect to get the same qualitative result.
Namely, that the maximal curvature is at the ends of the 
fourbranes. 
More interesting  is the case when the distance between the
D4-branes is smaller then $R_{10}$.
Consider first two semi-infinite D4-branes attached to
one NS brane from opposite sides.
Using eq.(\ref{lklk}) we find that the metric is
\be ds^2=dvd\bar{v}\left( 1+\frac{c^2
  R_{10}^2}{|v-c/2|^2|v+c/2|^2}\right),\ee
and the curvature is
\be R=8\frac{\mid v+c/2\mid ^2\mid v-c/2\mid ^2
\mid v\mid ^2c^2R_{10}^2}
{(\mid v+c/2\mid ^2\mid v-c/2\mid ^2+c^2R_{10}^2)^3}.\ee
Again far on the branes the curvature vanishes.
The vanishing of the curvature at $v=0$ is due to the parity symmetry
of the curve. 
When $c\gg R_{10}$ the maximal value of the curvature is 
$ R_{max}\approx 1/R^2_{10}$.
When $c\ll R_{10}$ the maximal value of the curvature is 
\be R_{max}\approx \frac{c^2}{R_{10}^4}\ll 1/R^2_{10}.\ee

If the D4-branes are at the same side
we can use eq.(\ref{rere}) to find that the metric is
\be ds^2=dvd\bar{v}(1+\frac{R^2\mid v\mid ^2}
{\mid v^2-c^2/4\mid ^2}),\ee
and the curvature is
\be R=-2R_{10}^2\frac{\mid v^2-c^2/4\mid ^2 \mid v^2+c^2/4\mid ^2}
{(\mid v^2-c^2/4\mid ^2+R_{10}^2\mid v\mid^2)^3}.\ee
Again when $c\gg R_{10}$ the maximal value of the curvature is of the
order of $1/R_{10}^2$. But unlike the previous case when $c\ll R_{10}$
the maximal value of the curvature is 
\be 
R_{max}\approx\frac{32R_{10}^2}{c^4}\gg\frac{1}{R_{10}^2}
\ee
at $v=0$.

Now let us consider two NS fivebranes which are connected 
by one D4-brane. The relevant curve is 
\be 
t^2+\frac{v}{a}t+1=0,
\ee
The metric is
\be 
ds^2=dvd\bar{v}(1+\frac{R^2_{10}}{\mid v^2-a^2\mid}),
\ee
and the curvature is 
\be 
R=2R_{10}^2\frac{\mid v\mid ^2}{(\mid v^2-a^2\mid +R_{10}^2)^3},
\ee
When $a \rightarrow 0$ the distance between the NS branes 
diverges (at finite v) and hence we find that the maximal curvature is,
$\approx 1/R_{10}^2$.
When $a\gg R_{10}$ the maximal curvature is 
$ R_{max}\approx 
\frac{a^2}{R_{10}^4}\gg \frac{1}{R_{10}^2}$
at $v=a$. Notice that  like in the previous case the curvature is not
bounded by $1/R_{10}^2$ but can be arbitrary large. 

To summarize, when the distances between the branes are 
much larger then $R_{10}$ the maximal 
curvature on $\Sigma $ is of the order of  $1/R_{10}^2$,
and hence diverges in the type $II_A$ limit.

However, when the distances between the branes are  
smaller then $R_{10}$ the curvature is not bounded by
$1/R_{10}^2$ and can be arbitrary large.

\end{document}